\documentclass[12pt]{article}
\usepackage{epsf}
\def\be{\begin{equation}}
\def\ee{\end{equation}}
\def\bea{\begin{eqnarray}}
\def\eea{\end{eqnarray}}

\begin{document}
\begin{titlepage}
\begin{center}
{\Large \bf William I. Fine Theoretical Physics Institute \\
University of Minnesota \\}
\end{center}
\vspace{0.2in}
\begin{flushright}
FTPI-MINN-15/38 \\
UMN-TH-3447/15 \\
August 2015 \\
\end{flushright}
\vspace{0.3in}
\begin{center}
{\Large \bf Formation of hidden-charm pentaquarks in photon-nucleon collisions. 
\\}
\vspace{0.2in}
{\bf V. Kubarovsky$^{1}$ and M.B. Voloshin$^{2,3,4}$  \\ }
$^1$Thomas Jefferson National Accelerator Facility, Newport News, VA 23606, USA \\
$^2$William I. Fine Theoretical Physics Institute, University of
Minnesota,\\ Minneapolis, MN 55455, USA \\
$^3$School of Physics and Astronomy, University of Minnesota, Minneapolis, MN 55455, USA \\ and \\
$^4$Institute of Theoretical and Experimental Physics, Moscow, 117218, Russia
\\[0.2in]

\end{center}

\vspace{0.2in}

\begin{abstract}
The cross section for formation in $\gamma + p$ collisions of the recently found hidden-charm pentaquark states $P_c(4380)$ and $P_c(4450)$ is discussed and estimated. The studies of these resonances in photon beam experiments can be complementary to those in the LHCb experiment setting, and may be more advantageous for measurement of their additional decay channels. It is pointed out that both the relative importance of such decays and the yield of the resonances in the $\gamma + p$ collisions are sensitive to the internal dynamics of the pentaquarks and can resolve between theoretical models.  Specific numerical estimates are discussed within a simple `baryocharmonium' model, where the the observed $P_c$ resonances are composites of $J/\psi$ and excited nucleon states with the quantum numbers of $N(1440)$ and $N(1520)$.

\end{abstract}
\end{titlepage}

The newly discovered~\cite{lhcb} baryonic peaks $P_c(4380)$ and $P_c(4450)$ in the $J/\psi \, p$ system emerging from the decays $\Lambda_b \to J/\psi \, p \, K^-$ extend to the baryonic sector the set of the known in the mesonic sector exotic multiquark hadrons containing a heavy quark-antiquark pair.  The internal dynamics of such hadrons is a subject of an intensive discussion in the literature and it is yet to be seen whether models developed for the mesonic states resonances can be applied to the new baryonic states, and where those latter states fall within the spectrum of the discussed models. The so far suggested interpretations of the $P_c$ peaks include molecular states~\cite{cllz,cclsz,rno,he} `made from' a charmed baryon and an (anti)charmed meson, pentaquarks containing tightly correlated diquarks~\cite{mpr,amnss}, or colored baryon-like and meson-like constituents~\cite{mm,lebed}, and a model~\cite{mo} where the peak $P_c(4450)$ is interpreted as a composite made of the charmonium state $\chi_{c1}$ and the proton. Finally, it has been also suggested that at least one of the peaks is not a resonance at all, but rather a kinematical singularity due to rescattering~\cite{gmwy,lwz,mikhasenko,amnss} in the decay $\Lambda_b \to J/\psi \, p \, K^-$.

Clearly, resolving between the models and clarifying the nature of the discovered hidden-charm pentaquark peaks, and possibly searching for similar peaks with other quantum numbers, requires further experimental studies. The purpose of the present paper is to estimate the yield of the novel baryonic channels in a medium energy photon beam on a proton target where the pentaquark peaks should appear in the $s$ channel at the photon energy around 10\,GeV. Such experiments can be advantageous for detailed studies of the production and decay properties of the pentaquark resonances in comparison with the LHCb environment.  The discussed yield is determined by the branching fraction $Br(P_c \to \gamma + p)$, and it will be shown here that this papameter can be expressed in terms of $Br(P_c \to J/\psi p)$ by a relation similar to a vector dominance for the $J/\psi$. Although such dominance cannot be justified as a general rule, in the situation at hand it can be applied due to arguments based on the heavy quark properties and special kinematics of the processes involved. As a result the peak cross section for $\gamma + p \to P_c \to J/\psi + p$, proportional to $\left [ Br(P_c \to J/\psi + p) \right ]^2$, can reach tens of nanobarns or more, if $Br(P_c \to J/\psi + p) \sim 10\%$. Such relatively large cross section may allow fairly detailed studies of the pentaquarks and a search for other similar states. In particular, it may be realistic to study the decays of the $P_c$ states into $J/\psi p \pi$ and $J/\psi p \pi \pi$. As will be argued below, such decays should be prominent, if the $P_c$ states are dominantly a baryocharmonium, i.e. a hadroquarkonium-type~\cite{mv07,dv} composite of $J/\psi$ and excited nucleon states similar to the known resonances $N(1440)$ and $N(1520)$. Such pattern of the decays of the $P_c$ resonances would disfavor the molecular models~\cite{cllz,cclsz,rno,he}, where one would expect the natural decay channels into a charmed hyperon and a meson, or from the $\chi_{c1} p$ complex model~\cite{mo}, where the expected dominant decay is $P_c(4450) \to \chi_{c1}+p$. Naturally, any observation of the $P_c$ peaks in the $\gamma p$ cross section would strongly disfavor the interpretation~\cite{gmwy,lwz,mikhasenko,amnss} in terms of `accidental' singularities in the $\Lambda_b$ decays.

For a resonance $P_c$ in the $s$ channel the cross section is given by the standard Breit-Wigner expression (see e.g. in Ref.~\cite{pdg}, Sec.~48.1.)
\be
\sigma(\gamma + p \to P_c \to J/\psi + p) = { 2 J +1 \over 4} \, {4 \pi \over k^2} \, {\Gamma^2/4 \over (E-E_0)^2 + \Gamma^2/4} \, Br(P_c \to \gamma + p) \, Br(P_c \to J/\psi +p)~,
\label{bw}
\ee
where $J$ is the spin of the $P_c$ resonance, $E=\sqrt{s}$ and $k$ is the center of mass (c.m.) momentum of the colliding particles. At the maximum of either of the $P_c$ resonances this expression gives numerically (at $k \approx 2.1\,GeV$)
\be
\sigma_{max}(\gamma + p \to P_c \to J/\psi + p) \approx { 2 J +1 \over 4} \, Br(P_c \to \gamma + p) \, Br(P_c \to J/\psi +p) \, 1.1 \times 10^{-27} {\rm cm}^2~.
\label{bwm}
\ee

The actual value of the cross section in Eq.(\ref{bwm}), naturally, depends on the product of the branching fractions, neither of which is presently known. However, it will be agrued here that the branching fraction $Br(P_c \to \gamma + p)$ can be estimated in terms of $Br(P_c \to J/\psi + p)$ in a way that does not directly rely on a specific model of the internal dynamics of the pentaquarks, but which is somewhat sensitive to the structure of the amplitude of the decay  $P_c \to J/\psi +p$. 

One can start with noticing that the coupling of hidden-charm pentaquark to the channel $\gamma + p$ is dominated by the electromagnetic current of the charmed quarks, $j_\mu = (\bar c \gamma_\mu c)$, so that for the purpose of the present evaluation of this coupling the light quarks can be considered electrically neutral. Indeed, the coupling of the photon to a light quark would additionally require a transition with annihilation of the $c \bar c$ pair into light hadrons which is very strongly suppressed for the $J/\psi$. 

The amplitude of the process $P_c \to \gamma + p$ can generally be written in terms of invariant form factors $F_\ell$ as
\be
A(P_c \to \gamma + p) = \sum_\ell F_\ell \, A_\ell~, 
\label{ff}
\ee
where $A_\ell$ are polarization-dependent structures whose number and specific form depends on the quantum numbers of the $P_c$ resonance. The form factors $F_\ell$ are Lorentz scalars and each is a function of three invariants: $F_\ell(P_\mu P^\mu ,p_\mu p^\mu,q^2)$, where $P$, $p$ and $q$ are the four-momenta of respectively the $P_c$ resonance, the proton, and the photon. At fixed $P_\mu P^\mu=M^2(P_c)$ and $p_\mu p^\mu = m_p^2$ each form factor is a real analytic function of complex variable $q^2$ with a dicontinuity at real positive $q^2$ corresponding to (hidden charm) intermediate states in the virtual photon channel. According to the dimensional constituent counting rule~\cite{mmt,bf} the form factors should fall off at large $|q^2|$ as $|q^2|^{-3}$, and should thus satisfy a dispersion relation without subtractions:
\be
F_\ell(M^2, m^2, q^2) = - {1 \over \pi} \, \int {{\rm Im}F_\ell (M^2,m^2,s) \over q^2 - s + i \epsilon} \, ds~.
\label{dr}
\ee
The intermediate states contributing to the dispersion integral are shown schematically in Fig.~1. 
  
\begin{figure}[ht]
\begin{center}
 \leavevmode
    \epsfxsize=16cm
    \epsfbox{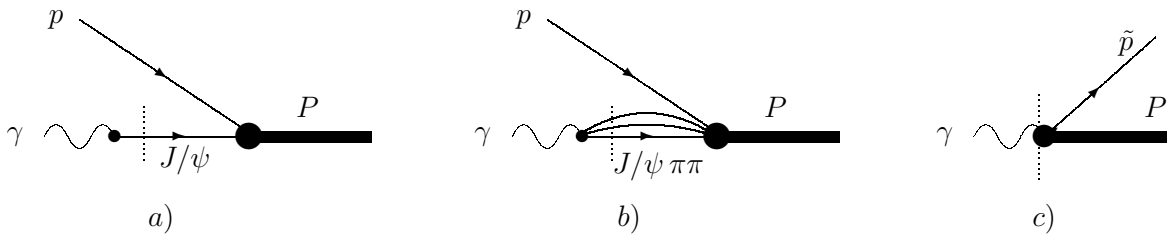}
    \caption{The types of contribution to the absorptive part of the amplitude $\gamma + p \to P_c$. The vertical dotted line indicates the unitary cut across an on-shell hidden-charm intermediate state.}
\end{center}
\end{figure} 

The graph $a)$ describes the contribution of the $J/\psi$ resonance and, as will be argued, gives the dominant part of the amplitude. The graph $b)$ corresponds to a continuum of intermediate states containing light mesons and charmonium with the invariant mass below $M-m \approx 3.44$\,GeV for $P_c(4380)$ and $M-m \approx 3.51$\,GeV for $P_c(4450)$. It is quite clear  that the only such state that is allowed by the quantum numbers and the kinematics is $J/\psi \, \pi \pi$. 
This contribution however should be exceedingly small. Indeed, the production of $J/\psi \, \pi \pi$ in $e^+e^-$ annihilation of this channel in the relevant range of invariant mass, i.e. in fact in the `tail' of the $J/\psi$ resonance, has never been observed experimentally. It is also expected to be very small theoretically~\cite{mv04} from a calculation based on the multipole expansion in QCD. Finally, the graph $c)$ corresponds to the `far cut' in the dispersion relation (\ref{dr}) and corresponds to the absorptive part in the amplitude $\gamma \to P_c \bar p + c.c.$. (This cut in the form factors $F_\ell(P_\mu P^\mu ,p_\mu p^\mu,q^2)$ appears at negative time components $p_0$ or $P_0$.) Such exotic process, also never observed in the $e^+e^-$ annihilation, can be expected to be very strongly suppressed by the form factor (apparently stronger than e.g. the form factor for the exclusive $p \bar p$ production in the $e^+e^-$ annihilation). In other words, the form factor, including its absorptive part, should be small 
at $q^2 > (M+m)^2$, and the contribution of the far cut in the integral in Eq.(\ref{dr}) can be neglected.

These considerations leave as significant only the contribution of the $J/\psi$ resonance shown in Fig.~1a, thus effectively reducing the calculation to a vector dominance relation between the form factors describing the amplitudes with a photon and the $J/\psi$ resonance, with the vertex for the $J/\psi - \gamma$ coupling readily determined from the leptonic width of $J/\psi$:  $\Gamma_{ee}(J/\psi) \approx 5.6\,$keV. It should be noted that such relation between the form factors is generally insufficient for relating the rates of the decays $P_c \to \gamma +p$ and $P_c \to J/\psi +p$ because of significantly different kinematics in these two processes. Indeed, the amplitude, as in Eq.(\ref{ff}), is generally contributed by partial waves with different orbital momentum $L$ with all values of $L$ being of the same parity, as determined by the spin-parity of the pentaquark. The c.m. momentum $p$ in the decays $P_c \to J/\psi +p$ is approximately 0.74\,GeV for $P_c(4380)$ and 0.81\,GeV for $P_c(4450)$, while the c.m. momentum $k$ in the decays $P_c \to \gamma + p$ is about 2.1\,GeV. Thus the contribution to the rate of a partial wave amplitude with orbital momentum $L$ is enhanced in the latter decay  by the factor $(k/p)^{2L+1}$ relative to its contribution in the former process. Specifically, if one writes the total width of the decay as a sum over $L$ of partial widths corresponding to different orbital waves: $\Gamma(P_c \to J/\psi + p) = \sum_L \, \Gamma_L(P_c \to J/\psi + p)$, the vector dominance relation gives
\be
\Gamma(P_c \to \gamma + p) = {3 \Gamma_{ee}(J/\psi) \over \alpha \, M(J/\psi)} \, \sum_L \,  f_L \, \left ( {k \over p} \right )^{2L+1} \,\Gamma_L(P_c \to J/\psi + p)~,
\label{vdr}
\ee
with $f_L$ standing for the fraction of the rate (in each partial wave) of the decay $P_c \to J/\psi +p$ that goes into transversally polarized $J/\psi$ resonance in the c.m. frame. The fraction $f_L$ is determined by the value of $L$ and the spin-parity of the pentaquark and is of order one. For instance, if $P_c(4380)$ has $J^P=(3/2)^-$ and $P_c(4450)$ has $J^P=(5/2)^+$ (one of the preferred by the data~\cite{lhcb} options), the allowed values of $L$ are 0 and 2 in the decay of $P_c(4380)$ and 1 and 3 in that of $P_c(4450)$. In this case one can readily find the corresponding values of the fractions $f_L$:
\newline
$P_c(4380):$
\be
f_0={2 \over 2+ \gamma^2} \approx 0.65~,~~~~~f_2={1 \over 1+ 2 \gamma^2} \approx 0.32~;
\label{f02}
\ee
$P_c(4450):$
\be
f_1={3 \over 3 + 2 \gamma^2} \approx 0.58~,~~~~~f_3={2 \over 2 + 3 \gamma^2} \approx 0.38~,
\label{f13}
\ee
where $\gamma$ stands for the gamma factor of the $J/\psi$ in the c.m. frame, $\gamma^2= 1+p^2/M^2(J/\psi)$.

Using this assignment of the quantum numbers for the observed pentaquarks and the relation (\ref{vdr}) of the width $\Gamma(P_c \to \gamma+p)$ to the partial widths of the decay $P_c \to J/\psi+p$, one arrives at the bounds for the formation cross section at the resonance maximum:
\bea
&& 1.5 \times 10^{-30}\, {\rm cm^2} \, < \, {\sigma_{max}[\gamma + p \to P_c(4380) \to J/\psi + p] \over Br^2[ P_c(4380) \to J/\psi+p]} \, < \, 47 \times 10^{-30} \, {\rm cm^2}~, \nonumber \\
&& 1.2 \times 10^{-29}\, {\rm cm^2} \, < \, {\sigma_{max}[\gamma + p \to P_c(4450) \to J/\psi + p] \over Br^2 [ P_c(4450) \to J/\psi+p]} \, < \, 36 \times 10^{-29} \, {\rm cm^2}~,
\label{modn}
\eea
where the lower bound corresponds to the presence of only the lower allowed partial wave, while the upper bound is found in the opposite situation where only the higher orbital wave is present. 

Currently neither the branching fraction $Br(P_c \to J/\psi +p)$ is known, nor the spin-parity quantum numbers for the observed pentaquarks are determined with certainty. For this reason it is not possible to estimate more definitely the value of the discussed formation cross section in a model independent way. For instance, in the model of Ref.~\cite{mo} the resonance $P_c(4450)$ is a $J^P=(3/2)^+$  state, while the peak $P_c(4380)$ is essentially ignored. Thus in this model the first line in Eq.(\ref{modn}) should in fact be applied to $P_c(4450)$, and the actual number for the cross section would depend on the unknown composition of the two partial waves in the decay. 

It is worth mentioning that a somewhat more definite prediction for the partial wave structure in the decays of {\em both} observed pentaquark states can be made in an alternative model of a baryocharmonium type, where the $P_c(4380)$ and $P_c(4450)$ resonances are made from the $J/\psi$ and excited nucleon states with respectively the quantum numbers of the known baryon resonances  $N(1440)$ and $N(1520)$. The spin-parity quantum numbers of the $N$-resonances combined with $J/\psi$ in an $S$-wave composite result in the assignment of the values of $J^P$ considered in the derivation of Eq.(\ref{modn}). Furthermore, in this case the conservation of the spin of $J/\psi$, required by the heavy quark spin symmetry, allows only the lowest partial wave in the decay of each of the resonances, so that the estimated cross section should be near the lower limit in the bounds (\ref{modn}). It can be also mentioned that the $N$ resonances are known to strongly decay into a nucleon and one or two pions (the latter includes some contribution from the decays going through $\Delta + \pi$). For this reason in such model one can expect that the decays of the pentaquarks $P_c \to J/\psi +p + \pi$ and $P_c \to J/\psi + p + \pi \pi$ should at least compete in the total rate with the observed $J/\psi+p$ channel, although it would be troublesome at present to assign any specific numbers to this expectation. Moreover, such baryocharmonium model has an obvious difficulty with understanding the masses of the pentaquarks. Indeed, although the mass splitting between $N(1520)$ and $N(1440)$ matches that between $P_c(4450)$ and $P_c(4380)$, the table values~\cite{pdg} of the masses themselves when combined with the mass of the $J/\psi$ exceed the experimental values by approximately 150\,MeV. It is by far not clear at present whether such a large required reduction in the mass of the composite due to the deformation of the excited nucleons by binding with $J/\psi$ is dynamically possible and thus whether such baryocharmonium model is viable. 

The initial experimental observation~\cite{lhcb} of the hidden-charm pentaquarks has posed a whole new slew of questions regarding such multiquark systems. It would be quite helpful for finding the answers if these and possible similar states could be produced and observed in experiments that would be additional to the current LHCb studies. In this paper the formation of the pentaquarks as $s$ channel resonances in $\gamma+p$ collisions is discussed and estimated as a possible setting for such experiments. It has been argued here that the amplitudes for the decays $P_c \to \gamma +p$ and $P_c \to J/\psi +p$ can be expressed through each other by a simple relation, formally coinciding with that of the vector dominance for $J/\psi$. Even though such dominance would not be applicable in general, in the discussed decays it is effectively enforced by the strong  suppression of nonresonant contributions in the dispersion relation in Eq.(\ref{dr}) and the absence of contribution of any other resonances besides $J/\psi$. Although the specific estimates of the observable cross section inevitably suffer from current uncertainties, amounting in fact to  more than two orders of magnitude spread in Eq.(\ref{modn}), even the lowest estimated values do not seem to be discouraging for the prospects of actually observing the pentaquark formation by a photon beam.

While this paper was finalized, the paper~\cite{wlz} appeared in the arXive also discussing the production and formation of the hidden-charm pentaquarks in $\gamma$ - nucleon collisions..

The work of M.B.V.  is supported in part by U.S. Department of Energy Grant No.\ DE-SC0011842. The work of V.K. is supported by the U.S. Department of Energy.  The Jefferson Science Associates (JSA) operates the Thomas Jefferson National Accelerator Facility for the United States Department of Energy under contract DE-AC05-06OR23177.


\begin{thebibliography}{99}
\bibitem{lhcb} 
  R.~Aaij {\it et al.} [LHCb Collaboration],
  arXiv:1507.03414 [hep-ex].

\bibitem{cllz} 
  R.~Chen, X.~Liu, X.~Q.~Li and S.~L.~Zhu,
  arXiv:1507.03704 [hep-ph].

\bibitem{cclsz} 
  H.~X.~Chen, W.~Chen, X.~Liu, T.~G.~Steele and S.~L.~Zhu,
  arXiv:1507.03717 [hep-ph].
  
\bibitem{rno} 
  L.~Roca, J.~Nieves and E.~Oset,
  arXiv:1507.04249 [hep-ph].
  
  
\bibitem{he} 
  J.~He,
  arXiv:1507.05200 [hep-ph].
  
\bibitem{mpr} 
  L.~Maiani, A.~D.~Polosa and V.~Riquer,
  arXiv:1507.04980 [hep-ph].
  
\bibitem{amnss} 
  V.~V.~Anisovich, M.~A.~Matveev, J.~Nyiri, A.~V.~Sarantsev and A.~N.~Semenova,
  arXiv:1507.07652 [hep-ph].
  
\bibitem{mm} 
  A.~Mironov and A.~Morozov,
  arXiv:1507.04694 [hep-ph].
  
\bibitem{lebed} 
  R.~F.~Lebed,
  arXiv:1507.05867 [hep-ph].
  
\bibitem{mo} 
  U.~G.~Mei{\ss}ner and J.~A.~Oller,
  arXiv:1507.07478 [hep-ph].
  
\bibitem{gmwy} 
  F.~K.~Guo, U.~G.~Mei{\ss}ner, W.~Wang and Z.~Yang,
  arXiv:1507.04950 [hep-ph].
  
\bibitem{lwz} 
  X.~H.~Liu, Q.~Wang and Q.~Zhao,
  arXiv:1507.05359 [hep-ph].
  
\bibitem{mikhasenko} 
  M.~Mikhasenko,
  arXiv:1507.06552 [hep-ph].
	
\bibitem{mv07} 
  M.~B.~Voloshin,
  Prog.\ Part.\ Nucl.\ Phys.\  {\bf 61}, 455 (2008)
  [arXiv:0711.4556 [hep-ph]].
	
\bibitem{dv} 
  S.~Dubynskiy and M.~B.~Voloshin,
  Phys.\ Lett.\ B {\bf 666}, 344 (2008)
  [arXiv:0803.2224 [hep-ph]].
	
\bibitem{pdg} 
  K.~A.~Olive {\it et al.} [Particle Data Group Collaboration],
  Chin.\ Phys.\ C {\bf 38}, 090001 (2014).
	
\bibitem{mmt} 
  V.~A.~Matveev, R.~M.~Muradian and A.~N.~Tavkhelidze,
  Lett.\ Nuovo Cim.\  {\bf 7}, 719 (1973).
\bibitem{bf} 
  S.~J.~Brodsky and G.~R.~Farrar,
  Phys.\ Rev.\ D {\bf 11}, 1309 (1975).
	
	
\bibitem{mv04} 
  M.~B.~Voloshin,
  Mod.\ Phys.\ Lett.\ A {\bf 19}, 665 (2004)
  [hep-ph/0402011].

	
\bibitem{wlz} 
  Q.~Wang, X.~H.~Liu and Q.~Zhao,
  arXiv:1508.00339 [hep-ph].

\end{thebibliography}
\end{document}